\documentclass[aps,prb,twocolumn,showpacs,superscriptaddress,amsmath,amssymb]{revtex4}
\usepackage{graphicx}
\usepackage{dcolumn}
\usepackage{bm}
\usepackage{color}

\begin{document}

\title{Quantum Hall states under conditions of vanishing Zeeman energy}

\author{F. J. Teran}
\affiliation{Laboratoire National des Champs Magn\'etiques Intenses, CNRS-UJF-UPS-INSA, 38042 Grenoble, France}
\affiliation{Instituto Madrile$\tilde{n}$o de Estudios Avanzados en Nanociencia, Ciudad Universitaria de Cantoblanco, 28049 Madrid, Spain}

\author{M. Potemski}
\affiliation{Laboratoire National des Champs Magn\'etiques Intenses, CNRS-UJF-UPS-INSA, 38042 Grenoble, France}

\author{D. K. Maude}
\affiliation{Laboratoire National des Champs Magn\'etiques Intenses, CNRS-UJF-UPS-INSA, 38042 Grenoble, France}

\author{T. Andrearczyk}
\affiliation{Institute of Physics, Polish Academy of Sciences, PL-02668 Warsaw, Poland}

\author{J. Jaroszynski}
\affiliation{Institute of Physics, Polish Academy of Sciences, PL-02668 Warsaw, Poland}
\affiliation{National High Magnetic Field Laboratory, Tallahassee, Florida 32306, USA}

\author{T. Wojtowicz}
\affiliation{Institute of Physics, Polish Academy of Sciences, PL-02668 Warsaw, Poland}

\author{G. Karczewski}
\affiliation{Institute of Physics, Polish Academy of Sciences, PL-02668 Warsaw, Poland}

\date{\today }

\begin{abstract}
We report on magneto-transport measurements of a two-dimensional electron gas 
confined in a Cd$_{0.997}$Mn$_{0.003}$Te quantum well structure under 
conditions of vanishing Zeeman energy. The electron Zeeman energy has been 
tuned via the $s-d$ exchange interaction in order to probe different quantum 
Hall states associated with metallic and insulating phases. We have observed 
that reducing Zeeman energy to zero does not necessary imply the disappearing 
of quantum  Hall states, \emph{i.e.} a closing of the spin gap. The spin gap value 
under vanishing Zeeman energy conditions is shown to be dependent on the filling 
factor. Numerical simulations support a qualitative description of the experimental 
data presented in terms of a crossing or an avoided-crossing of spin split Landau 
levels with same orbital quantum number $N$.
\end{abstract}

 \maketitle

\section{Introduction}

In a quantizing magnetic field, the physics of a two-dimensional electron gas 
(2DEG) at very low temperatures is controlled by electron-electron interactions 
which dominate over the single particle physics leading to exotic collective ground states. For example, the spin exchange energy often dominates over the single particle Zeeman energy in the fractional and integer quantum Hall regimes.\cite{Cho,Nicholas,Usher,Koch,Schme,Dane,Jungwirth,Leadley,dePoor,Jaro,Vakili,Freire,Fischer,Desrat,Wie,Maude,Girvin,Bychkov,Kallin,Jungwirth2} The rich physics of the quantum Hall
ferromagnetism has attracted considerable interest when two spin split Landau levels with different spin index (\emph {i.e.}$\sigma$) and orbital quantum number($N$) are brought in coincidence.\cite{Nicholas,Koch,Dane,Jungwirth,dePoor,Vakili,Jaro,Desrat} In this case, the quantum Hall effect (QHE) may exhibit hysteretic features with a complicated phenomenology\cite{Jaro,dePoor} related to phase transitions at non-zero temperatures. Such typical behavior of Ising ferromagnetic systems(\emph{i.e.} with easy-axis anisotropy)\cite{Jungwirth} can be explained in terms of domain walls \cite{Jungwirth} or distinct symmetry-broken states\cite{Freire}.

Equally, the crossing of spin split Landau levels with the same $N$ and different spin index leads to interesting physics\cite{Usher,Maude,Leadley}. In this case, QHE exhibits a different phenomenology related to no finite-temperature phase transitions without hysteretic features. Such typical behavior of Heisenberg ferromagnetic systems (\emph{i.e.} isotropic anisotropy)\cite{Jungwirth} leads to the existence of a non-zero electron spin gap in the energy spectrum of a 2DEG under conditions of vanishing Zeeman energy. This behavior has been observed in GaAs-based structures in the integer quantum Hall regime at $\nu=1$ and $3$. This could be explained as a consequence of the well established model\cite{Bychkov,Kallin} of electron-electron exchange interaction (at least for $\nu=1$) which forces a ferromagnetic ordering of the electronic spins at odd filling factors.\cite{Usher, Leadley,Maude} Therefore, any small perturbation opens the electron spin gap ($\Delta_{S}$) even when Zeeman energy is zero. Reasoning in these terms also implies that the exchange contribution to $\Delta_{S}$ always dominates over the Zeeman term. This is in agreement with routine experimental observations in 2DEGs confined in III/V based structures which clearly show that $\Delta_{S}$ deduced from activated transport data at odd filling factors always greatly exceeds the single particle Zeeman energy.\cite{Usher,Leadley,Maude}. In addition, electronic spins cannot be treated as an isolated system but unavoidably interact with localized spins of nuclei\cite{Barret} or localized spins of substitutional magnetic ions, such us Mn$^{2+}$.\cite{Jaro,Teran,Knobel,Buhmann,Gui} The incorporation of localized magnetic moments in diluted magnetic semiconductors (DMS) offers unique spin splitting engineering\cite{Wojto} via the ${s-d}$ exchange interaction between the electron and magnetic ion spins\cite{Furdyna}. This can potentially be exploited to probe the spontaneous ferromagnetic order of a 2DEG via $\Delta_{S}$ when the electron Zeeman energy is continuously tuned through zero via the $s-d$ exchange interaction with localized spins.

\begin{figure}
\includegraphics[width=0.9\linewidth,angle=0,clip]{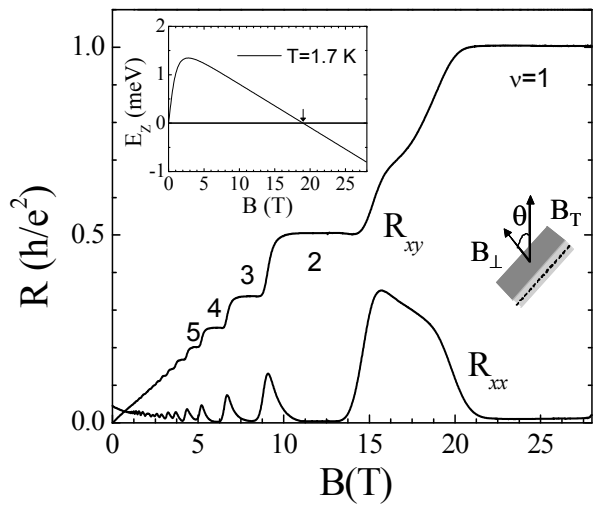}
\caption{ (Color on line) Magnetic field dependence of $R_{xx}$ and $R_{xy}$ under perpendicular magnetic fields at $T=1.7$~K for our sample with an electron density $n_{e}=5.9 \times 10^{11}$~cm$^{ - 2}$. Left inset: Magnetic field dependence of $E_{Z}$ at $T=1.7$~K calculated according to Eq.\ref{eq1}. Right inset: Schematic diagram indicating sample's orientation to magnetic field.}\label{fig1}
\end{figure}

Here, we report on the study of QHE in a n-type modulation doped Cd$_{0.997}$Mn$_{0.003}$Te quantum well (QW) structure under conditions of vanishing Zeeman energy. So far, the standard method to tune the electron g-factor through zero in semiconductor structures was to apply hydrostatic pressure\cite{Maude}. However, this method offers rather limited range of tunability of the Zeeman energy and can also modify the carrier density or the mobility. The incorporation of substitutional magnetic moments such as Mn$^{2+}$ in CdTe QW structures offers an additional possibility of tuning the electron spin splitting via the ${s-d}$ exchange interaction between the spins of 2DEG and the localized magnetic moments\cite{Teran3}. In a simple approach, the extended 2D electron and localised Mn$^{2+}$ spins can be treated as two paramagnetic subsystems. Thus, a single electron spin feels the external magnetic field and the mean field resulting from the Mn$^{2+}$ spin polarization. Due to the opposite signs of the bare Zeeman and ${s-d}$ exchange terms in a Cd$_{1-x}$Mn$_{x}$Te QW, the magnetic field ($B$) dependence of the total electron Zeeman energy ($E_{Z}$) results from the competition between these two contributions. At low magnetic fields where the ${s-d}$ exchange interaction term dominates, $E_{Z}$ has large and positive values related to the ${s-d}$ exchange term. On increasing $B$, the bare Zeeman term increases linearly whereas the ${s-d}$ exchange interaction term saturates. Therefore, at high magnetic fields $E_{Z}$ decreases because of the negative contribution related to the bare Zeeman term (see left inset of Figure \ref{fig1}). Thus, $E_{Z}$ reaches zero value at a given magnetic field for which Zeeman term is equal to ${s-d}$ exchange term what depends on the Mn concentration. Thus, the particular $B$-dependence of $E_{Z}$ in Cd$_{1-x}$Mn$_{x}$Te QW structures allows to study the evolution of a given quantum Hall state when $E_{Z}$ is continuously tuned from positive to negative values through zero. This can be achieved by simply tilting the plane of the 2D electron sheet with respect to the $B$-direction. That offers an unique  opportunity to probe the ferromagnetic phenomenology of Heisenberg systems when the value of $E_{Z}$ is modulated and the sign is inverted. The clear advantage of DMS systems is the wide range over which the g-factor can be continuously tuned without any changes in the mobility or carrier density. The disadvantage lies in the lower mobility (below 200.000 cm$^{2}$V$^{-1}$s$^{-1}$) of a 2DEG confined in magnetic QWs\cite{Kolk} with respect to their III-V non-magnetic counterparts. The experimental results presented in this work show that the condition of vanishing effective Zeeman energy in Cd$_{1-x}$Mn$_{x}$Te QWs does not necessarily imply that $\Delta_{S}$ is zero. We also observe that the opening or closing of the $\Delta_{S}$ when $E_{Z}=0$ tightly depend on the filling factor.

This paper is organized as follows. In Sec. II, we present a description of the sample structure and the experimental procedure. In Sec. III, we show the magneto-transport data obtained from the tilted-field experiments, notably the evolution of longitudinal ($R_{xx}$) and transversal ($R_{xy}$) resistances as a function of $E_{Z}$ and temperature for different filling factors. The $R_{xy}$ shows the development of plateau between $\nu =2$ and $1$. This resistance feature is not associated with any quantum Hall state but rather related to the condition $E_{Z}=0$. Moreover, the evolution of $R_{xx}$ minima as the Zeeman energy passes through zero indicates that $\Delta_{S}$ remains open at $\nu=3$but closes at $\nu=5$. In addition, the evolution of the $R_{xx}$ maxima on either side of $\nu=3$ (around $\nu=7/2$ and $5/2$) under $E_{Z}=0$ conditions is markedly different, suggesting a different opening of $\Delta_{S}$. In Sec. IV, we present numerical simulations to qualitatively describe and discuss the $\Delta_{S}$ dependence on $E_{Z}$ and $B$ at distinct filling factors: $\nu=3$ and $5$ and $\nu=7/2$, $5/2$ and $3/2$. Reasoning in terms of the density of states ($DOS$) at the Fermi level ($E_{F}$), we find two different situations which lead to distinct amount of $DOS$ at the Fermi level ($DOS(E_{F})$) under conditions of vanishing Zeeman energy. This depends upon whether the $E_{F}$ lies inside either the upper or the lower spin level of a given spin split Landau level. Finally in Sec. V, we summarize the important points found in this work.

\section{Experimental details}

For the investigation, an n-type modulation doped structure made of a 10 nm thick Cd$_{1-x}$Mn$_{x}$Te QW embedded between Cd$_{0.8}$Mg$_{0.2}$Te barriers was employed. The effective Mn concentration in the QW is $x=0.003$ determined from low magnetic field transport measurements\cite{Teran}. The 2DEG is formed by placing iodine doped n-type Cd$_{0.8}$Mg$_{0.2}$Te layer separated from the Cd$_{0.997}$Mn$_{0.003}$Te QW by $10$~nm thick undoped Cd$_{0.8}$Mg$_{0.2}$Te spacer. For the magneto-transport measurements a standard mesa etched Hall-bar was defined (typical dimensions $0.5 \times 1$~mm). The 2D electron concentration obtained from either the Hall resistance or the periodicity of the
Shubnikov de Haas oscillations is n$_{e}=5.9 \times 10^{11}$~cm$^{-2}$ and the mobility $\mu_{e}=60.000$~cm$^{2}$V$^{-1}$s$^{-1}$ at liquid helium temperatures. The sample was mounted on a rotation stage to tilt the sample with respect to magnetic field direction at temperatures ($T$) down to $50$~mK and magnetic fields up to $28$~T. The sample resistance components were measured using \emph{ac} techniques ($10.7$~Hz with $I \sim 10 - 100$~nA). Whereas $E_{Z}$ is defined by the total magnetic field ($B_{T})$, the filling factor is determined by the magnetic field component which is perpendicular to the 2D plane ($B_{\bot})$ (see right inset in Fig. \ref{fig1}). Thus, by  rotating the sample the value of $E_{Z}$ at a given filling factor is varied from positive to negative values passing through zero. 

\begin{figure}[tbp]
\includegraphics[width=1\linewidth,angle=0,clip]{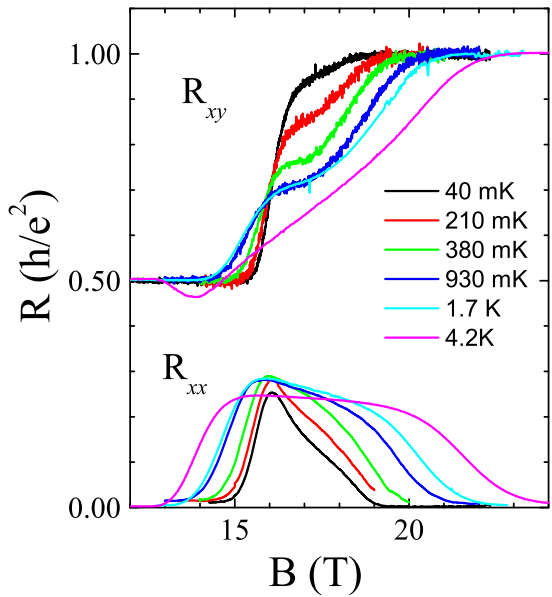}
\caption{(Color on line) Temperature dependence of $R_{xx}$ and $R_{xy}$ in the vicinity of $\nu=3/2$.}\label{fig2}
\end{figure}

\section{Magneto-transport results}
\subsection{Longitudinal and transverse resistance at perpendicular
magnetic fields}

Figure~\ref{fig1} shows the magneto-transport measurements of our sample under perpendicular magnetic fields up to $B=28$~T at $T=1.7$~K. A well developed Landau quantization at the integer quantum Hall regime is observed: well pronounced Shubnikov-de Haas oscillations in $R_{xx}$ and quantized plateaus in $R_{xy}$. The rather broad $R_{xy}$
plateaus associated to low filling factors (for instance, $\nu=1$ and $2$) and the low 2DEG mobility reveal the presence of relevant disorder in our sample. At low $B$ (where $E_{Z}\geq \hbar\omega_{c}$), Shubnikov de Haas oscillations show complex beating patterns which have been discussed in detail in our previous work.\cite{Teran} At higher $B$ (where $\hbar\omega_{c} > $ $E_{Z}$), both $R_{xx}$ and $R_{xy}$ show a typical QHE for a spin polarized 2DEG: resistance plateaus and minima associated with a continuous sequence of odd and even integer filling factors. However, certain $R_{xx}$ and $R_{xy}$ irregularities appear in the range of $B=14-22$~T. At these magnetic fields, the $R_{xx}$ maxima between $\nu=2$ and $1$ is anomalously distorted so that the development of an additional minimum in $R_{xx}$ could be envisaged around $B=17$ T. In parallel, a kink, resembling the development of quantum Hall plateau, is observed in the $R_{xy}$ trace. These anomalies have been previously related to the appearance of the $\nu=4/3$ fractional quantum Hall state\cite{Teran2}. However, the $T$-dependence of $R_{xy}$ and $R_{xy}$, shown in Figure \ref{fig2}, rules out such a possibility. While there is some evidence for the existence of a plateau at intermediate temperatures (ranging 210 mK - 1.7 K), the feature washes out at the lowest temperatures rather than fully developing as expected for a fractional state. Indeed, the value of the anomalous plateau tightly depends on temperature: whereas at the lowest temperature ($T$=40 mK), the $R_{xy}$ anomalous plateau almost reaches the value of $h/e^{2}$ , the  value drops to 0.5$h/e^{2}$ at the highest $T$ (4.2K). The non-quantization of the developing plateau is a strong indication that an alternative explanation to fractional states should be sought. Simultaneously, the $R_{xx}$ maxima broadens when increasing temperature. In order to clarify if the vanishing of $E_{Z}$ is related to these resistance anomalies, we model the $B$ and $T$-dependence of $E_{Z}$ in our sample by assuming that:

\begin{equation}\label{eq1}
E_{Z} = g_{e}\mu_{B} B + E_{s-d}\mathbb{B}_{5/2}(B,T,T_{0})
\end{equation}

where $g_{e}=-1.6$ is the electron g-factor in CdTe\cite{Sirenko}, $\mu _{B}$ is the Bohr magneton, $E_{s-d}=1.25$~meV is the $s-d$ exchange interaction term between free electron and localized Mn$^{2+}$ determined here from the low magnetic field transport measurements \cite{Teran}, $\mathbb{B}_{5/2}(B,T,T_0)$ is the modified Brillouin function, and $T_{0}$ is the effective temperature of the Mn spin subsystem\cite{Furdyna} considering the magnetization correction due to the antiferromagnetic Mn$^{2+}$-Mn$^{2+}$ interactions. In our particular sample, $T_{0}=0.12$~K was determined from transport measurements at low magnetic fields.\cite{Teran} The expected $B$-dependence of $E_{Z}$ in our sample at $T=1.7$~K is shown in the inset of Fig.\ref{fig1}. At low magnetic fields, $E_{Z}$ has large and positive values related to the exchange term. Once the exchange term is saturated, $E_{Z}$ decreases linearly with magnetic field and crosses zero around $B \sim 19$~T, \emph{i.e.}, in the magnetic field range in which the $R_{xx}$ and $R_{xy}$ anomalies are observed. 

\begin{figure}[tbp]
\includegraphics[width=0.8\linewidth,angle=0,clip]{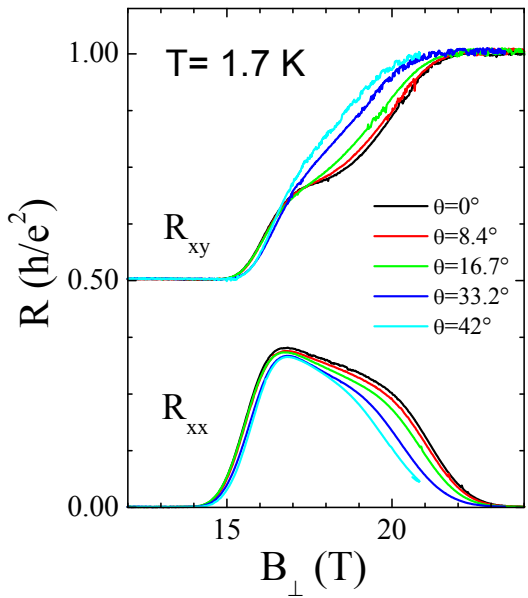}
\caption{(Color on line) $R_{xx}$ and $R_{xy}$ as a function of perpendicular magnetic fields, $B_{\perp}$, for different tilted angles $\theta$ in the vicinity of $\nu=3/2$.}\label{fig3}
\end{figure}

In order to probe the relationship between the $R_{xy}$ and $R_{xx}$ anomalies associated to $\nu=3/2$ and the $E_{Z}=0$ condition, we have performed tilted-field magneto-transport experiments shown in Figure~\ref{fig3}. The resistance anomalies observed at $B=19$~T weaken when tilting the sample, \emph{i.e.} when the $E_{Z}=0$ condition is shifted towards lower $B_{\perp}$: the $R_{xx}$ high field shoulder decreases and the $R_{xy}$ non-quantized plateau significantly increases its value. This strongly suggests that those resistance anomalies observed between $B=14-24$~T are related to the vanishing of $E_{Z}$ which occurs in this magnetic field range as shown at the left inset of Fig.\ref{fig1}.

To further elucidate the influence of a vanishing Zeeman energy on the magneto-transport properties, we have investigated the influence of tuning the Zeeman energy on integer quantum Hall states. At integer filling factors and low temperatures, our 2DEG shows zero resistivity and zero conductivity ($\rho_{xx}=\sigma _{xx}=0$) indicating the existence of a well developed gap. Characteristically for insulating systems, the gap can be determined by thermally activated transport measurements. This procedure yields reliable gap values for clean 2DEG\cite{Leadley} (\emph{i.e.} narrow Landau level broadening). However, this is not the case for our 2DEG as revealed by the wide $R_{xy}$ plateaus  observed in Fig.\ref{fig1} and the low carrier mobility. Such plateau broadening can be related to the intrinsic disorder associated with ternary compounds, to the long range disorder associated with iodine donor impurities and/or with the presence of background dopants in the QW region. In a first approach, all these disorder mechanisms increase Landau level broadening leading to a large number of localized states. Nevertheless, thermally activated gaps obtained by $T$-dependence measurements in our sample may provide a qualitative evolution of $\Delta_{S}$ when tuning $E_{Z}$ at given filling factor. Thus, we investigate the evolution of $R_{xx}$ when Zeeman energy vanishes at $\nu=3$ and $5$ (insulating-like quantum Hall states). Furthermore, we similarly study the evolution of $R_{xx}$ when the vanishing Zeeman energy is brought into coincidence with $\nu=7/2$, $5/2$, and $3/2$ (metallic-like quantum Hall states).

\begin{figure}[tbp]
\includegraphics[width=1\linewidth,angle=0,clip]{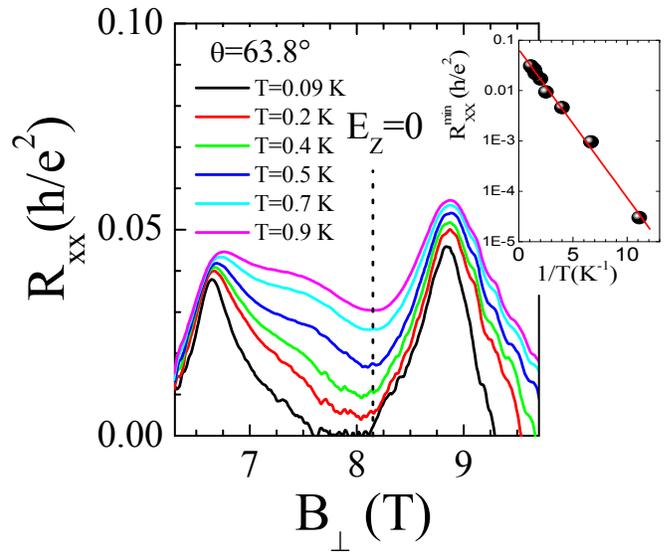}
\caption{(Color on line) Temperature dependence of $R_{xx}$ minima related to $\nu=3$ for a tilted angle $\theta =63.8^\circ$ for which $E_{Z}=0$ is expected at $B_{\bot0}=8.1$~T. Inset: Arrenhius plot of the $T$-dependence of $R_{xx}$ at $B_{\bot0}=8.1$~T.}\label{fig4}
\end{figure}

\subsection{Spin gaps at $\nu=3$ and $5$ versus Zeeman splitting}

Figure~\ref{fig4} shows the $R_{xx}$ measured in the $T$-range between 90 and 900 mK and at a fixed tilted angle $\theta=63.8^\circ$. At this particular angle, vanishing Zeeman energy (\emph{i.e.} $E_{Z}=0$) and $\nu=3$ coincide at $B_{\perp}$=8.1 T. Under this condition, a well defined $R_{xx}$ minimum is observed, reaching zero value below $200$~mK. This clearly indicates that $\Delta_{S}$ at $\nu=3$ remains open even under vanishing Zeeman energy conditions. Assuming a thermally activated behavior, the measured value of the resistance minima ($R_{xx}^{min}$) at $\nu=3$ can be described by the following expression:

\begin{equation}  \label{eq2}
R_{xx}^{min} = R_{0}\cdot \exp( -E_A/2kT)
\end{equation}

a plot of $ln(R_{xx}(T))$ versus $1/T$ allows us to deduce an activation energy $E_A=0.25$~meV with the pre-factor $R_{0} =0.1 h/e^{2}$ (see inset of Fig.\ref{fig4}). In case of a dirty 2DEG, $E_A$ should be increased by the width of the region of extended states ($\Gamma_{ext}$) in order to obtain a better estimate of spin gap\cite{Leadley}. Thus, $\Delta_{S}$=$E_A$+$\sqrt{2}\Gamma_{ext}$. Simulations of $R_{xx}$ at low fields were performed to model the magneto-transport measurements for our 2DEG system\cite{Teran} considering $\Gamma_{ext}=0.2$~meV. This value of $\Gamma_{ext}$ is in agreement with the value of Landau level broadening ($\Gamma_{ext}$=0.075 meV) employed in numerical simulations described in Section IV. Hence, we obtain a value of $\Delta_{S}=0.53$~meV at $\nu=3$ under conditions of vanishing Zeeman energy.

\begin{figure}[tbp]
\includegraphics[width=1\linewidth,angle=0,clip]{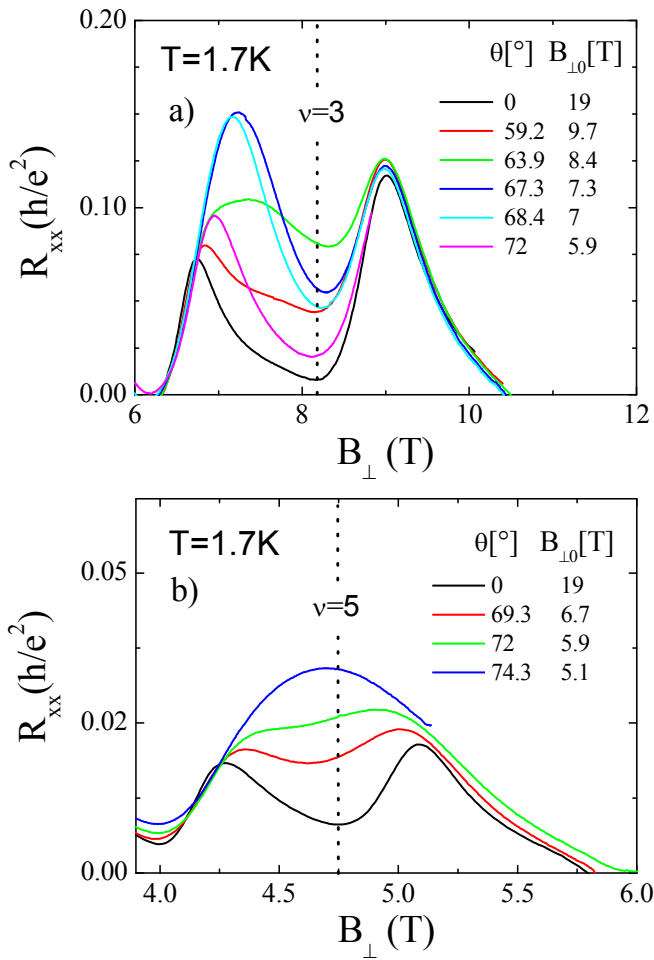}
\caption{(Color on line) (a) $R_{xx}$ as a function of perpendicular magnetic field $B_{\perp}$ for different tilt angles $\theta$ in the vicinity of $\nu=3$ at $T=1.7$~K. (b) $R_{xx}$ as a function of perpendicular magnetic field $B_{\perp}$ for different tilt angles $\theta$ in the vicinity of $\nu=5$ at $T=1.7$~K. The perpendicular magnetic field ($B_{\bot 0}$) for which $E_{Z}=0$ is indicated for each angle.}\label{fig5}
\end{figure}

Figure~\ref{fig5}(a) shows the evolution of $R_{xx}$ when the $E_{Z}=0$ condition is tuned through $\nu=3$ at a fixed temperature. At $\theta=0^\circ$, $R_{xx}$ shows a well developed minima at $\nu=3$ which becomes less pronounced when tilting the sample. At $\theta=63.9^\circ$, the conditions $E_{Z}=0$ and $\nu=3$ coincide. At that angle, the value of $R_{xx}$ at $\nu=3$ shows a maximum. In contrast, $R_{xx}$ at $\nu=3$ recovers minimum values at larger angles (see data at $\theta= 72^\circ$). From Eq.\ref{eq1}, we obtain the expected perpendicular magnetic field ($B_{\perp 0}$) for which $E_{Z}=0$ at a given angle as indicated in the Figure ~\ref{fig5}. As expected, the $R_{xx}$ minimum associated with $\nu=3$ becomes less pronounced as $E_{Z}$ is initially decreased when the sample is rotated away from the normal configuration. In contrast, the resistance minimum at $\nu=3$ shows its maximum value when $E_{Z}\approx 0$. Afterwards, the resistance minimum at $\nu=3$ decreases again for large tilt angles when $E_{Z}$ large and negative values.  

\begin{figure}[tbp]
\includegraphics[width=0.8\linewidth,angle=0,clip]{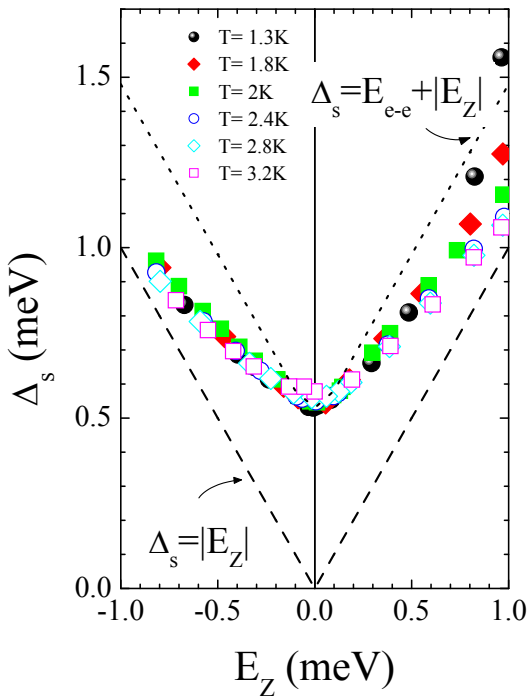}
\caption{(Color on line) Spin gap ($\Delta _{S}$) as a function of Zeeman energy at $\nu=3$ for different temperatures. $\Delta _{S}$ was obtained by using Eq.(\ref{eq2}) as explained in the text. The dotted line represents the spin gap dependence $\Delta _{S}=E_{Z}+E_{e-e}$ and the dashed line $\Delta _{S}=E_{Z}$.}\label{fig6}
\end{figure}

The results of similar measurements around $\nu=5$ are quite different (see Fig.\ref{fig5}(b)). The $R_{xx}$ minimum is transformed into a maximum under vanishing Zeeman energy conditions. This indicates that, at least for the measurement temperature employed ($T=1.7$~K), $\Delta_{S}$ collapses under conditions of vanishing Zeeman energy. Characteristically, a perfect overlap for the $N=2$ spin split Landau levels when $E_{Z}\sim 0$ transforms the $R_{xx}$ minima  at $\nu=5$ into a maximum whose amplitude is approximately twice that of the adjacent resistance maxima at $\nu=11/2$ and $9/2$. The factor of two can be roughly understood assuming that $R_{xx}$ is proportional to $DOS(E_{F})$\cite{Burgt}, which indeed approximately doubles when two spin Landau levels overlap. Finally, we have not found any evidence of hysteretic resistance traces around $\nu=3$ and $5$ quantum Hall states at the temperature studied ($T=1.7$~K). Assuming a thermally activated behavior from Eq.(\ref{eq1}), the measured value of $R_{xx}$ minima at $\nu=3$ for a fixed temperature (Fig.\ref{fig5}(a)) can be used to determine $E_A$ as follows: 

\begin{equation}  \label{eq3}
{E}_{A} ({E}_{Z}) = 2k{T}( - ln({R}_{xx}^{min} ) + ln({R}_{0}))
\end{equation}

where $R_{0} = 0.1 h/e^{2}$ obtained from Arrhenius plot under $E_{Z}=0$ at $\nu=3$ conditions as mentioned above. The value of $E_A$ was determined using Eq.(\ref{eq2}), for a large number of angles (\emph{i.e.} different $E_{Z}$ values at $\nu=3$) and six different temperatures ranging between $1.3$ and $3.2$~K. Figure \ref{fig6} shows $\Delta_{S}=E_A+\sqrt{2}\Gamma_{ext}$ as a function of $E_{Z}$. The values of $\Delta_{S}$ determined from data taken at different temperatures all collapse onto the same curve which validates our assumption of a thermally activated behavior. As expected, $\Delta _{S}$ has a minimum value, $\Delta_{S}=0.53$~meV, for a tilt angle $\theta \approx 64^\circ$ where $E_{Z}=0$ in agreement with the $\Delta _{S}$ value obtained from fitting Eq.\ref{eq1} shown in the inset of Fig.\ref{fig4}. Such a $\Delta_{S}$ value under vanishing Zeeman energy conditions seems extremely small in comparison to values of spin gap enhanced by electron-electron interactions. As mentioned above, the large Landau level broadening associated to this dirty 2DEG may lead to imprecise gap values from thermally activated measurements in the studied temperature range. However,it is enough to provide a qualitative description of the $\Delta_{S}$ dependence on $E_{Z}$. Thus, Figure~\ref{fig6} indicates that at $\nu=3$, $\Delta_{S}\geq E_{Z}$ and clearly remains open when $E_{Z}=0$. Either side of $E_{Z}=0$, $\Delta_{S}$ increases linearly with $E_{Z}$ but with different slopes, slightly steeper in the positive $E_{Z}$ range than in the negative-one. Such specific character of the $\Delta_{S}$ slopes may be related to the different behavior of $R_{xx}$ maxima either sides of $\nu=3$ when $E_{Z}$ is tuned (see Fig.\ref{fig5}(a)). Inspecting Fig.\ref{fig6}, we notice that $\Delta _{S} \geq  |E_{Z}|$ when $| E_{Z}|$ is sufficiently large. On the other hand, $\Delta_{S}>> | E_{Z}|$ only when $E_{Z} \approx 0$. Reasoning in terms of spin split Landau levels, such characteristic dependence can be related to the effect of resonant level repulsion. We assign this phenomenon to electron-electron interaction related to electron spin polarization at $\nu=3$. However as mentioned above, it is important to note that the opening of $\Delta_{S}$ in terms of resonant repulsion (avoided crossing) of spin split $N=1$ Landau levels is not exactly what one would expect from the accepted knowledge of the physics of a 2DEG at odd filling factors in high quality III-V nanostructures. For a disorder free two-dimensional electrons, we would expect $\Delta_{S}$ to be the sum of the Zeeman energy and the electron-electron exchange energy, $E_{e-e}$, which is constant for a given filling factor and independent of the Zeeman energy. This would imply that $\Delta_{S}=E_{e-e}+E_{Z}$ which is not reflected in our experimental data (see dotted line in Fig.\ref{fig6}. A model of spin Landau level repulsion (\emph{i.e.} avoided crossing) is shown to successfully explain our observation as will be described in Section IV.

We summarize this section by pointing out the contrasting observation of the avoided crossing of spin Landau levels at $\nu=3$ while a crossing of spin Landau levels is observed at $\nu=5$.

\subsection{Vanishing Zeeman splitting at half integer
filling factors ($\nu=7/2$, $5/2$ and $3/2$)}

We now turn our attention to the behavior of $R_{xx}$ under vanishing Zeeman energy conditions when $E_{F}$ lies in the center of a Landau level, \emph{i.e.} metallic states, at $\nu=7/2$ ($B_{\perp}\sim 6.6$~T), $\nu=5/2$ ($B_{\perp}\sim 9$~T) and $\nu=3/2$ ($B_{\perp} \sim 19$~T). The traces of $R_{xx}$ at $T=1.7$~K and different tilt angles shown in Fig.\ref{fig5}(a) allow us to follow the evolution of the $R_{xx}$ maxima at $\nu=5/2$ and $7/2$ when $E_{Z}$ is tuned through zero in the range $+1$ to $-1$~meV. The behavior of the $\nu=5/2$ and $7/2$ $R_{xx}$ peaks is significantly different. The position and amplitude of the resistance maximum associated with $\nu =5/2$ remains independent of the effective Zeeman energy. In contrast, $R_{xx}$ maximum associated with $\nu=7/2$ is strongly influenced by $E_{Z}$. As $E_{Z}$ is decreased at $\nu=7/2$ by tilting the sample, the $\nu=7/2$ $R_{xx}$ maxima broadens (see data measured at $\theta= 59.2^\circ$) on the high magnetic field side and its amplitude increases. At $\theta = 67.3^\circ$, the sample is under conditions of $E_{Z}=0$ at $\nu=7/2$. In this situation, the amplitude of the $R_{xx}$ maximum related to $\nu=7/2$ is roughly twice the value at $\theta= 0^\circ$. At larger tilt angles, $E_{Z}$ is increasingly negative leading to the recovery of $R_{xx}$ trace observed at $\theta=0^\circ$. Assuming that $R_{xx}$ is proportional to the $DOS(E_{F}$), it is intuitively clear that the doubling of the $R_{xx}$ maximum can be interpreted as a result of a double degeneracy (perfect overlap) of spin-split $N=1$ Landau levels. We consider that the effect of the $R_{xx}$ increase can be seen as a characteristic feature indicating some overlap of the otherwise spin split $N=1$ Landau levels. An unchanged resistance would correspond to a situation in which the $DOS(E_{F})$ remains unchanged and therefore $\Delta_{S}$ remains open, which is the case observed for the $\nu=5/2$ $R_{xx}$ maximum. In terms of our model of ``repulsing levels'', we deduce that spin split $N=1$ Landau levels may cross at $\nu=7/2$while they avoid to cross at $\nu=5/2$. We have found no evidence of hysteretic resistance traces around $\nu=7/2$ and $5/2$ quantum Hall states at the temperature range studied. 

Turning back to the tilted-field behavior of the $R_{xx}$ peak around $\nu=3/2$, we note that this case is a mixing of the $\nu=5/2$ and $7/2$ cases. As shown in Fig.\ref{fig3}, the amplitude of the $R_{xx}$ peak observed around $\nu=3/2$ is essentially not affected by tilting the sample as in case of $\nu=5/2$. In contrast, the high field side of $R_{xx}$ peak narrows when tilting the sample as shown for $\nu=7/2$ large angles ($\theta > 67.3^\circ$) where $E_{Z}$ increases. Simultaneously, the $R_{xy}$ plateau disappears. This particular behavior can be interpreted as the result of a slight overlap between the spin split $N=0$ Landau levels at $B_{\perp} \sim 19 T$ which diminishes by tilting the sample (\emph{i.e.} increasing $E_{Z}$). The slight overlap can be interpreted as an avoided crossing of spin split $N=0$ Landau levels under vanishing Zeeman energy conditions. Numerical simulations presented in next Section IV will qualitatively support this interpretation.

\begin{figure}[tbp]
\includegraphics[width=1\linewidth,angle=0,clip]{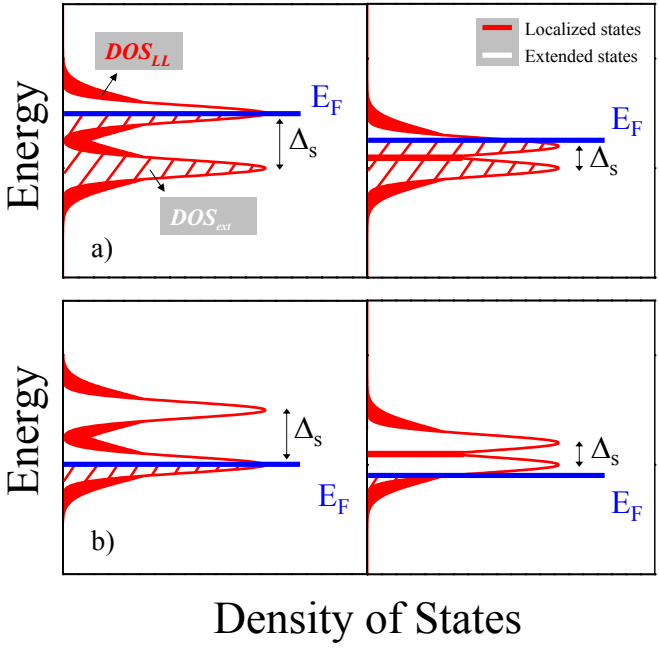}
\caption{(Color on line) Schematic illustration of the Gaussian $DOS_{LL}$ and $DOS_{ext}$with widths $\Gamma_{LL}$ and $\Gamma_{ext}$, respectively, in two different conditions of Landau level mixing (right side: mixing, left side: no mixing) when $E_{F}$ lies  a) in spin-up Landau level (as in case of $\nu=3/2$ and $7/2$) and b) spin-down Landau level (as in case of $\nu=5/2$).}\label{fig7}
\end{figure}

We summarize this section by pointing out the observation of the avoided crossing of spin Landau levels at $\nu=5/2$ and $3/2$ in contrast with the observed crossing of spin Landau levels at $\nu=7/2$.

\begin{figure}[tbp]
\includegraphics[width=1\linewidth,angle=0,clip]{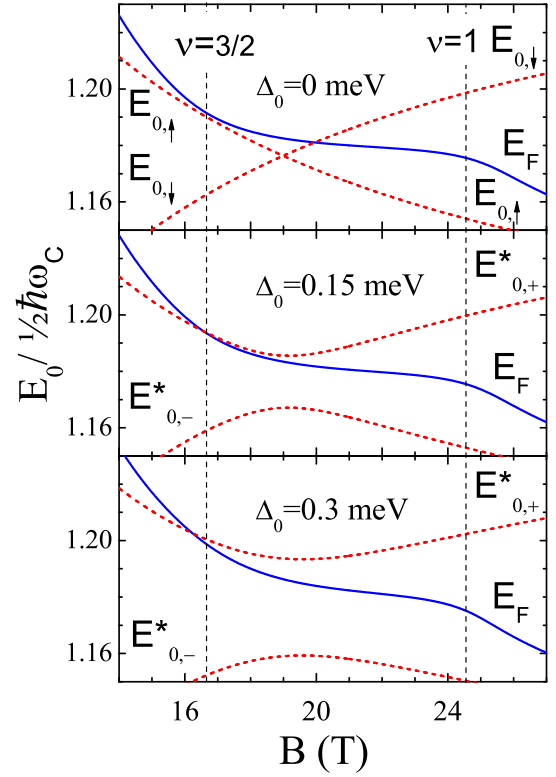}
\caption{(Color on line) The calculated $E_{F}$(solid line) and energy of the spin split $N=0$ Landau levels normalised to 1/2$\hbar\omega_{c}$(dotted lines) as a function of magnetic field for different values of $\Delta _{0}$ and $T=1.7$~K.}\label{fig8}
\end{figure}

\section{Numerical simulations}

In order to qualitatively account the directly visible consequences of different $\Delta_{S}$ values on the magneto-transport properties of the 2DEG, we have performed numerical calculations. In the following section, we develop a simple model which phenomenologically introduces the spin splitting in terms of an energy of interaction for the avoided crossing of the spin split Landau levels. The model is capable of qualitatively reproducing the experimental data and gives some insight into the evolution of $\Delta_{S}$ with $E_{Z}$ and filling factor. In order to do that, we assume different interaction terms depending on filling factor. In addition, we assume that the total density of the electronic states under magnetic field can be described by a set of Gaussian broadened Landau levels. The energy of an electron under quantizing magnetic fields which interacts with Mn$^{2+}$ ions within a single particle picture is given by:

\begin{equation}  \label{eq4}
E_{{N},\sigma}= ({N}+1/2) \hbar \omega_{c} + {E}_{Z}\sigma 
\end{equation}

where $N=0,1,2,...$ corresponds to the Landau level index, $\sigma$ denotes the electron angular momentum whose z-component has the value +1/2(-1/2) for spin up(down), $\hbar\omega_{c}=eB_{\bot}/m^{\ast}_{e}$ denotes the cyclotron energy and m$^{\ast}_{e}= 0.107$~m$_{o}$ is the electron effective mass determined by cyclotron resonance\cite{Sadowski}. $E_{Z}$ is given by Eq.\ref{eq1}. In addition, the $DOS$
of spin resolved Landau levels with Gaussian broadening is described as

\begin{equation}  \label{eq5}
{DOS}_{LL} = \Sigma_{N,\sigma}\frac{e{B}}{h}\frac{1}{\Gamma_{LL} \sqrt{2\pi}} \exp \left( -{\frac{%
\left( {E - {E}_{{N},\sigma} } \right)^2}{2\Gamma_{LL} ^2}} \right)
\end{equation}

where $\Gamma_{LL}$ is the Landau level Gaussian broadening parameter. We neglect the effective thermal distribution when calculating $E_{F}$ assuming $T=0$~K for electrons. The procedure of calculating $R_{xx}$ and $R_{xy}$ distinguishes between localized and extended states\cite{Burgt} as shown in Fig.7. The calculation procedure considers that only the latter ones contribute to the conductivity. In order to do that, we assume that in the central part of each Landau level there exists a band of extended states of Gaussian form whose $DOS$ is given by 

\begin{equation}  \label{eq6}
{DOS}_{ext} = \Sigma_{N,\sigma} \frac{e{B}}{h}\frac{1}{\Gamma_{ext} \sqrt{2\pi} }\exp \left( -{\frac{%
\left( {E - {E}_{N} \uparrow \downarrow } \right)^2}{2\Gamma_{ext}^2 }} \right)
\end{equation}

where $\Gamma_{ext} < \Gamma_{LL}$. For a given perpendicular magnetic field, $E_{F}$ is calculated numerically by integrating over the total density of states\cite{Burgt}(\emph{i.e.} $DOS_{LL}$). $R_{xx}$ is then taken to be proportional to the density of extended states at $E_{F}$, \emph{i.e.} $DOS_{ext}(E_{F})$ and $R_{xy}$ is proportional to the integral of $R_{xx}$.\cite{Burgt} Our experimental data show that the spin gap remains open even under vanishing $E_{Z}$ conditions. In order to reproduce such a behavior, we assume that spin split Landau levels with index $N$, $E_{N \uparrow , \downarrow }$,  are replaced by interacting levels $E^{*}_{N +,-}$ to calculate $DOS_{LL}$ and $DOS_{ext}$. The energy levels $E^{*}_{N +,-}$ for $N$=0 are given by

\begin{equation}  \label{eq7}
{E}^{*}_{0 + , - }= \frac{1}{2} ({E}_{{0} \uparrow } - {E}_{{0}
\downarrow } )\pm \{\frac{1}{4}( {E}_{{0} \uparrow } - {E}_{{0} \downarrow })^{2} +(%
\frac{\Delta _{0}}{2})^{2})\}^{1/2}
\end{equation}

where $\Delta _{0}$/2 is the energy of interaction for the avoided crossing. Physically, $\Delta _{0}$ corresponds to the experimentally observed spin gap $\Delta _{S}$ under vanishing Zeeman energy conditions (\emph{i.e.} when $E_{Z}=0$).

The adjustable parameters in our simulations are; the total width of the Landau level ($\Gamma_{LL}$), the width of the band of the extended states ($\Gamma_{ext}$) and the interaction parameter $\Delta_{0}$. We have found that our data are best described when $\Gamma_{LL}=0.15$~meV and $\Gamma_{ext}=0.075$~meV both assumed to be constant in the magnetic field range considered ($B=4-28$~T). The value of $\Gamma_{LL}$ corresponds to a scattering time $\tau=2.75 \times 10^{-11}$~s in reasonable agreement with the transport scattering time $\tau _{t} \approx 3.65 \times 10^{-11}$ s deduced from the sample mobility. On the other hand, the interaction parameter $\Delta _{0}$ was found to be dependent on magnetic field as will be described below.

\begin{figure}[tbp]
\includegraphics[width=1\linewidth,angle=0,clip]{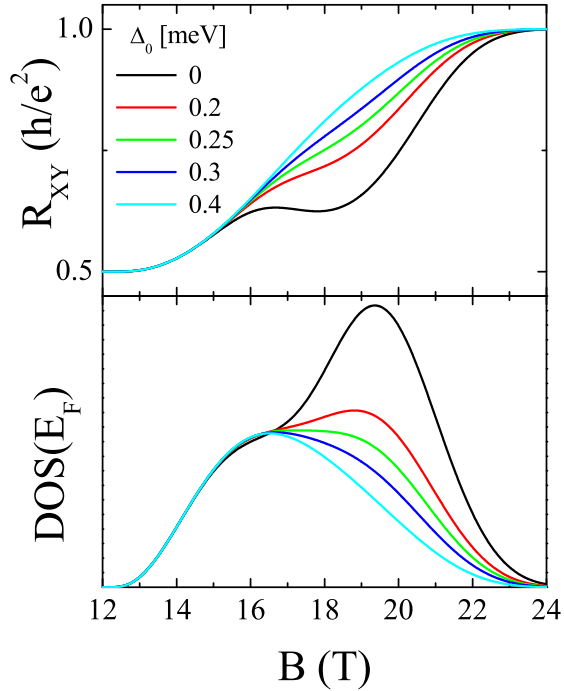}
\caption{(Color on line)Calculated $R_{xy}$ and $DOS(E_{F})$ as a function of magnetic field for different values of $\Delta _{0}$ and $T=1.7$~K.}\label{fig9}
\end{figure}

\subsection{Vanishing spin gaps at half integer filling factors ($\nu=7/2$, $5/2$ and $3/2$)}

Initially, we focus our attention on modeling the resistance anomalies observed near $B\sim 19$~T at $\nu=3/2$ (see Fig.\ref{fig2}). The non-quantized $R_{xy}$ plateau and the high-field shoulder of $R_{xx}$ can be understood due to the floating up of $E_{F}$ when the spin split $N=0$ Landau levels approach under vanishing Zeeman energy conditions. This is tightly related to $DOS(E_{F})$. The description of $DOS(E_{F})$ depends on the value of $\Delta_{0}$ as can be seen in Fig.\ref{fig8}. There, we represent traces of the calculated evolution the spin split $N=0$ Landau levels and $E_{F}$ as a function of magnetic field for different values of $\Delta_{0}$ and $T=1.7$~K (required to calculate $E_{Z}$). When $\Delta_{0}=0$~meV, the spin split $N=0$ Landau levels cross at $B \sim 19$~T. $E_{F}$ goes slightly above the crossing of spin split $N=0$ Landau levels leading to an additional peak of the $DOS_{ext}(E_{F})$ and a pronounced broadening as  shown in Figure \ref{fig9}; at the same time a kink appears in $R_{xy}$ which resembles a plateau development (see in Fig.\ref{fig9}). When increasing $\Delta_{0}$ value till $0.2$~meV, the $DOS(E_{F})$ reduces when  $E_{F}$ goes through the center of the upper spin branch Landau level (see Fig.\ref{fig8}) over a wide range of magnetic fields. The $E_{F}$ behavior with respect to spin split Landau levels gives rise to broad and pronounced $DOS_{ext}(E_{F})$ peak around $B=19$~T but with a smaller amplitude than in case of $\Delta_{0}=0$ meV. Simultaneously, the $R_{xy}$ kink increases its resistance value around $B=19$~T with respect to $\Delta_{0}=0$ meV (see Fig.\ref{fig9}). In contrast, for larger $\Delta_{0}$ values ($0.25-0.4$~meV), the spin split $N=0$ Landau levels have a more pronounced energetic distance leading to a significant reduction of $DOS(E_{F})$ when $E_{F}$ goes from $E_{0,+}$ towards $E_{0,-}$. The  increase of magnetic field raises the degeneracy of the Landau levels pulling down $E_{F}$with respect to the center of the upper spin branch (see Fig.\ref{fig8}). This is reflected in the transport properties as a transformation of the $DOS_{ext}(E_{F})$ peak into a high field shoulder which becomes less pronounced as $\Delta_{0}$ further increases (see Fig.\ref{fig9}. Simultaneously, the $R_{xy}$ kink continuously increases its resistance value and become less pronounced when increasing $\Delta_{0}$(see Fig.\ref{fig9}). 

\begin{figure}[tbp]
\includegraphics[width=1\linewidth,angle=0,clip]{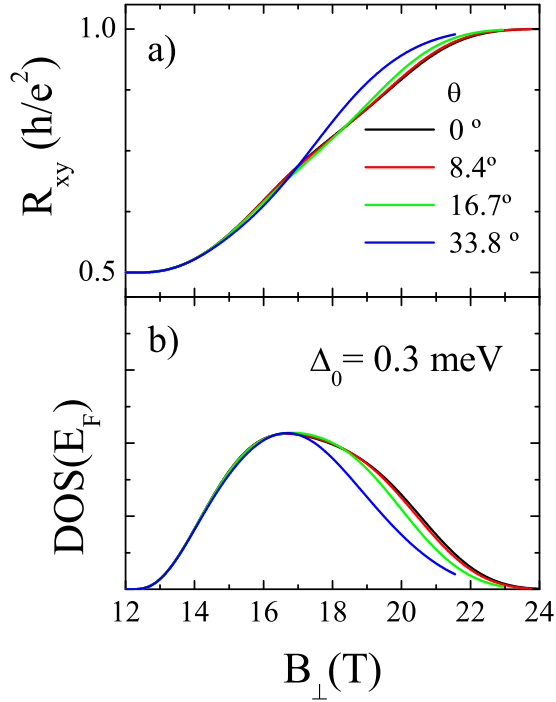}
\caption{(Color on line)Calculated $R_{xy}$ and $DOS(E_{F})$ as a function of magnetic field for different tilt angles in the vicinity of $\nu=3/2$ for $\Delta _{0} = 0.3$~meV and $T=1.7$~K.}\label{fig10}
\end{figure}

In addition, the effect of tilting the sample can be reasonably modeled for $\nu=3/2$ by considering $\Delta _{0}=0.3$~meV. As shown in Figure \ref{fig10}, the simulation qualitatively reproduces our observations fairly well (compare with Fig.\ref{fig3}): the value of the $R_{xy}$ plateau increases with the tilt angle $\theta$, reaching the value of the adjacent integer quantum Hall plateau (\emph{i.e.} $h/e^{2}$) for $\theta > 33^\circ$. At this titled angle, the $R_{xy}$ trace shows an unique slope while for smaller angles two different slopes can be appreciated, in good agreement with the experimental data (see \ref{fig3}). Simultaneously, the $R_{xx}$ maximum become narrower exclusively on the high field side when tilting the sample. At the same time the shoulder shifts to lower $B_{ \bot }$ and becomes less pronounced, qualitatively in good agreement with experiments. In our simulations, the behavior of $E_{F}$ depends on a single parameter which is the ratio of the interaction energy to the Landau level broadening ($\Delta _{0}$/$\Gamma_{LL}$) so that the determined value of $\Delta_{0}$ depends on the Landau level broadening used. 
Another important limitation of the model corresponds to the description of the $T$-dependence of the plateau. In a first approach, we may expect that increasing temperature the number of localized states should decrease.\cite{Rigal} This may therefore affect the plateau formation and also the broadening of the $R_{xx}$ minima associated with $\nu=1$ as we clearly see in Fig.\ref{fig2}. In our calculations, changing $\Delta_{0}$ should lead to similar effects which is not the case (see Fig.\ref{fig9}).

Our model qualitatively describes the $R_{xy}$ anomaly as the result of an avoided crossing of spin split $N=0$ Landau levels which slightly overlap under vanishing Zeeman energy conditions for a given $\Delta _{0}$/$\Gamma_{LL}$ ratio. In the standard picture of the integer QHE, the quantization of Hall plateaus is assigned to electron localization. Moving away from integer filling factors, the electron density and their localization length increases. Hence, free electrons become delocalized leading to the increase of $R_{xy}$ because the plateau quantization is lost. Thus, we may expect a non-quantized plateau when Landau level cross under vanishing Zeeman energy condition. This particular circumstance related to the mixing of localized-delocalized states at the Fermi level when $E_{Z}=0$ (see Fig.\ref{fig7}) originates the reentrant of an insulating phase responsible of the observation of a non-quantized plateau. Under such spin Landau level mixing conditions, the thermal activation determines the value of the $R_{xy}$ as shown in Fig.\ref{fig2}. Reasoning in similar terms, this model also explains the appearance of the $R_{xx}$ high field shoulder at $\nu=3/2$. The success of our qualitative model represents an important goal which allows to predict the appearance of non-quantized plateaus each time Landau level crossing and $E_{F}$ coincide. The important parameters which favor the observation of such resistance anomalies are $\Delta _{0}$ and $\Gamma_{LL}$. The disorder favors the observations of such resistance phenomena: high-magnetic field $R_{xx}$ shoulder (instead of well pronounced peaks) and well developed non-quantized $R_{xy}$ plateau (instead of Delta function-like plateau reductions \cite{Freire}). Moreover, our observations reveals that, at the temperature range studied, the crossing (or avoided crossing) of Landau levels with the same $N$ and different $\sigma$ index does not lead to hysteretic resistance behaviour, contrary to the case with different $N$ and $\sigma$ indexes.\cite{Jaro,Freire}

\begin{figure}[tbp]
\includegraphics[width=1\linewidth,angle=0,clip]{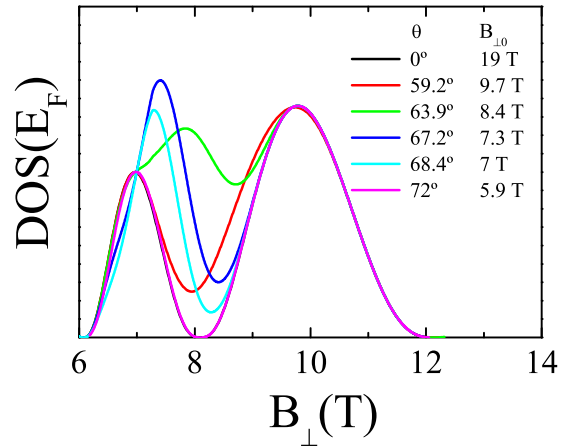}
\caption{(Color on line) Calculated $DOS_{ext}(E_{F})$ as a function of magnetic field in the vicinity of $\nu=3$ considering $\Delta _{0} = f(B_{\bot })$ and $T= 1.7$~K as described in the text. The perpendicular magnetic field for which $E_{Z}=0$ ($B_{\bot 0}$) is noted for each angle.}\label{fig11} 
\end{figure}

Although we conclude that spin Landau levels avoid to cross in both the $\nu=3/2$ and $\nu=5/2$ case, the corresponding $R_{xx}$ resistance maxima behave quite differently when $E_{Z}$ is varied: high field shoulder disappearing at $\nu=3/2$ and no changes at $\nu=5/2$. We think, this ''asymmetry'' results from the fact that partial overlap of spin Landau levels affects the position of $E_{F}$ in a different way depending on whether $E_{F}$ is located in the vicinity of the upper or of the lower spin level. Our reasoning is schematically illustrated in Fig.\ref{fig7}. The situation when $E_{F}$ is in the vicinity of the upper spin level, which, for example, corresponds to the case of a 2DEG at $\nu=3/2$, is considered in Fig.\ref{fig7}(a). In Fig.\ref{fig7}(b), $E_{F}$ is in the vicinity of the lower spin level which, for example, illustrates the $\nu=5/2$ case. In the left side panel of Fig.\ref{fig7} we consider well separated spin levels and the condition at which the maximum in the $R_{xx}$ should occur ($E_{F}$ located exactly at the center of the corresponding Landau level). Let us now imagine that we increase the magnetic field but at the same time the spin levels approach leading to a partial overlap. The primary effect of the increase of the magnetic field is to increase the Landau level degeneracy and therefore to lower $E_{F}$ with respect to the center of the corresponding spin Landau level. This simply translates into a decrease of the corresponding resistance by assuming that $R_{xx} \sim DOS(E_{F}$). However, as shown in Fig.\ref{fig7}, the position of $E_{F}$ (with respect to the center of the corresponding spin Landau level) is also affected by the increasing overlap of the spin levels. We note that the energy level overlap leads to the decrease of $DOS$ available below the center of the upper spin level and to the increase of $DOS$ below the center of the lower level. Therefore, increasing the overlap between spin levels we push $E_{F}$ up if it is in the vicinity of the upper spin level and push it down if it is in the vicinity of the lower spin branch. We expect that the additional ``overlap induced'' downwards shift of $E_{F}$ may somehow speed up the decay of the corresponding $R_{xx}$ peak, but the effect of ``overlap induced upwards shift'' of $E_{F}$ may have much more pronounced consequences on the observed shape of the corresponding $R_{xx}$ maxima. 

Therefore, we may qualitatively understand the effect of the broadening of the $\nu=3/2$ resistance maximum observed on the high field side when separation between spin Landau levels becomes small. Even more severe distortions of this resistance peak could be expected, for example, the situation when $E_{F}$ crosses twice the center of the same spin level and therefore the observation of the splitting of the $\nu=3/2$ peak should be, in principle, possible. We show in the next section that our qualitative considerations are confirmed by a more quantitative approach to simulate numerically our experimental results.

\begin{figure}[tbp]
\includegraphics[width=0.8\linewidth,angle=0,clip]{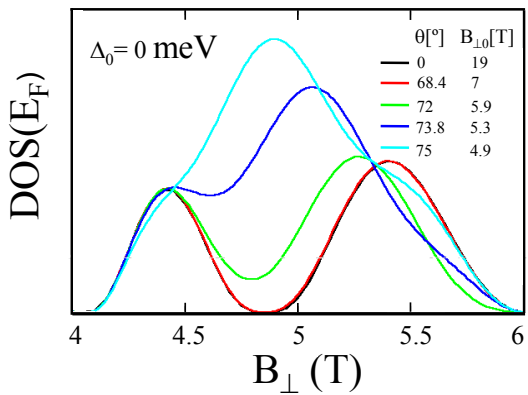}
\caption{(Color on line) Calculated extended $DOS(E_{F})$ as a function of magnetic field in the vicinity of $\protect\nu $=5 considering $\Delta _{0}=0$~meV and $T= 1.7$~K as described in the text. The perpendicular magnetic field for which $E_{Z}=0$ ($B_{\bot 0}$) is noted for each plot.}\label{fig12}
\end{figure}

\subsection{Spin gaps at $\nu=3$ and $5$ versus Zeeman splitting}

We now turn to the discussion of the shape of the $R_{xx}$ traces obtained as the effective Zeeman energy is swept through zero in the range $4 < \nu < 2$ shown in Fig.\ref{fig5}(a). We know that the gap at $\nu=3$ and $\nu=5/2$ remains open with a $\Delta _{0}=0.3$~meV. The doubling of the resistance peak at $\nu=7/2$ under conditions of
vanishing Zeeman energy suggests that $\Delta_{S}$ closes (\emph{i.e.} $\Delta _{0} \ll \Gamma_{LL}$). It is clear that there is a strong filling factor dependence of $\Delta _{0}$ which rapidly changes from zero to finite values. A rapid opening of $\Delta_{S}$ is well known in GaAs based 2DEG's and is generally explained in terms of the exchange energy which is proportional to the characteristic Coulomb energy and proportional to the spin polarization of the system. Since, for broadened Landau levels, the spin polarization depends on the size of $\Delta_{S}$ a positive feedback occurs and the onset of spin splitting is generally very abrupt. This can be modeled but requires that the gap and spin polarization be calculated self-consistently. We make the simplistic approximation that the residual gap depends only on the Coulomb energy so that $\Delta _{0} \propto \sqrt B $. The calculated $DOS_{ext}(E_{F})$ as a function of perpendicular magnetic field for different tilt angles and $T=1.7$~K (required to calculate $E_{Z}$) are shown in Fig.\ref{fig11}. Given the simplicity of the model used the agreement with the measured resistance (see Fig.\ref{fig5}(a)) is reasonably good. As $E_{Z}$ passes through zero the peak around $\nu \approx 7/2$ increases in amplitude and shifts to higher fields. To complete our simulations, we consider $R_{xx}$ around $\nu=5$ and show that it is possible to roughly reproduce the experimental results (see Fig.5.(b)). Assuming $\Delta_{0}=0$ and $T=1.7$~K, our numerical simulations support the qualitative description of the experimental data concluding that spin Landau levels cross (\emph{i.e.} perfect spin Landau level overlap) when $E_{Z}=0$ leading to $R_{xx}$ maxima, as shown in Figure \ref{fig12}.

\section{Summary}
We have studied the evolution of quantum Hall states under conditions of vanishing Zeeman energy. The experimental results presented here show that the condition of vanishing effective Zeeman energy does not necessarily imply the closing of the spin gap. We observe that the opening or closing of the spin gap is a function of the filling factor. Thus, $\Delta _{S}$ remains closed for $\nu=5$ and $7/2$ and opened for $\nu=3, 5/2$ and $3/2$ at the temperature studied (T=1.7 K). Moreover, the experimental procedure allows the investigation of quantum Hall states under vanishing Zeeman energy conditions not only for insulating quantum Hall states (\emph{i.e.} $\nu=3$ and $5$) but also for metallic states in the vicinity of $\nu=3/2$, $5/2$ and $7/2$. A phenomenological model based on an avoided crossing between the two spin split components of the corresponding Landau level qualitatively reproduce the $R_{xx}$ and $R_{xy}$ measurements at different conditions. Thus, we show that the mixing of localized and delocalized states under avoided Landau level crossing conditions leads to the appearance of a non-quantized plateau in the $R_{xy}$. Our observations can be explained assuming that $\Delta _{S}$ evolves in terms of e-e interactions mediated by disorder leading to a different filling factor dependence.

\section{Acknowledgments}
This work has been supported by the Foundation for Polish Science through subsidy 12/2007, the European Union through the EuroMagNET contract n$^\circ$~228043 and the Innovative Economy grant POIG.01.01.02-00-008/08 related to the European Regional Development Fund.

\end{document}